\documentclass{aastex}

\newcommand{\etal}{\it{et al.}\rm}

\slugcomment{Submitted to the Astronomical Journal}
\shorttitle{The Asymmetric Thick Disk}
\shortauthors{Parker, J.E., Humphreys, R.M., \etal}

\begin{document}

\title{The Asymmetric Thick Disk: A Star Count and Kinematic Analysis. II The Kinematics}

\author{Jennifer E. Parker\altaffilmark{1},  Roberta M. Humphreys\altaffilmark{1}} \affil{Department of Astronomy, 
University of Minnesota, Minneapolis, MN 55455} \email{parker@astro.umn.edu}
\email{roberta@aps.umn.edu}

\and

\author{Timothy C. Beers} \affil{Department of Physics and Astronomy, 
Michigan State University, East Lansing, MI 48824}
\email{beers@pa.msu.edu}

\altaffiltext{1}{Visiting Astronomer, Kitt Peak National Observatory and Cerro Tololo 
Inter-American Observatory, National Optical Astronomy 
Observatory, which is operated by the Association of Universities for Research in 
Astronomy, Inc. (AURA) under cooperative agreement with the National Science 
Foundation.}

\begin{abstract}

We report a  kinematic signature associated with the observed asymmetry in the 
distribution of thick disk/inner halo stars interior to the Solar circle  described in 
Paper I. In that paper we found a statistically significant excess (20 -- 25\%)
of stars in quadrant I ({\it l} $\sim$ 20$\arcdeg$ to 55$\arcdeg$) both above and below 
the plane ({\it b} $\sim$ $\pm$ 25$\arcdeg$ to $\pm$ 45$\arcdeg$) compared to the 
complementary region in quadrant IV. We have measured Doppler velocities for 741 stars, 
selected according to the same magnitude and color criteria, in the direction of the asymmetry 
and in the corresponding fields in quadrant IV. We have also determined spectral types and 
metallicities measured from the same spectra.  We not only find an asymmetric 
distribution in the V$_{LSR}$ velocities for the
stars in the two regions, but the angular rate of rotation ($\omega$) for
the stars in quadrant I reveals a  slower effective rotation rate compared to the 
corresponding quadrant IV stars. The results for  V$_{LSR}$ and $\omega$ also show 
an interesting  dependence on Galactic longitude which is most pronounced in quadrant I.   
We use our [Fe/H] measurements to separate the stars into the three primary population 
groups, halo, thick disk, and disk, and conclude that it is primarily the thick disk stars  
that show the slower rotation in quadrant I. These stars  are also responsible for the 
observed variation of V$_{LSR}$ and $\omega$ with Galactic longitude. A solution for the radial,
tangential and vertical components of the V$_{LSR}$ velocities, reveals  a significant lag of
$\approx$ 80 to 90 km s$^{-1}$ in the direction of Galactic rotation for the thick disk stars in 
quadrant I, while in quadrant IV, the same population has only a $\sim$ 20 km s$^{-1}$ lag 
confirming the kinematic asymmetry between the two directions. In Paper I we  
concluded that the asymmetry in the star counts could be best explained by either a 
triaxial thick disk or an interaction between the bar in the disk and the thick disk/inner 
halo stars. The results reported here support a rotational lag among the thick disk stars 
due to a gravitational interaction with the bar as the most likely explanation for the asymmetry
in both the star counts and the kinematics. The affected thick disk stars, however, may be associated
with the recently discovered Canis Major debris stream or a similar merger event.

\end{abstract}


\section{Introduction}

In \citet{ph03}, hereafter Paper I, we reported evidence for an asymmetrical
excess of thick disk/inner halo stars extending from approximately {\it l}
$\sim$ 20$\arcdeg$ to 55$\arcdeg$ and {\it b} $\sim$ 25$\arcdeg$ to 45$\arcdeg$ both 
above and below the plane. In Paper I we considered three possible
explanations for the asymmetry, including a possible merger remnant, but concluded 
that either a triaxial thick disk \citep{bs91},  with its major axis in quadrant I or an interaction 
with the bar in the disk \citep{her92,dbs98} best explained the observations. Both of these would explain 
the star count excess but with different kinematics. A triaxial thick disk/inner halo, 
for example, may have a distinct rotational velocity about the Galactic center different from the disk. 
Indeed thick disk stars are observed to have a lag of 30-50 km s$^{-1}$ \citep{rei98,cb00} with 
respect to disk stars.  
However a fast rotating bar in the disk in quadrant I with co-rotation 
at 3--4 kpc from the Galactic center could induce a gravitational ``wake'' that would 
trap and pile up stars behind it \citep{her92,dbs98}. In response to the bar, 
there would not only be an excess of stars, but some of the stars might show a measurable 
``lag'' or slower rotation as a result. Thus in either case, if the asymmetry in the star
counts is due to a structural feature, then it should be supported by an asymmetrical 
pattern in the stars' motions. To search for a kinematic signature associated with the asymmetry in the
star counts that could also  provide additional evidence for the cause of the asymmetry, 
we have obtained medium resolution spectra for radial velocities, spectral classification,
and metallicity estimates of stars selected from 12 fields, 6 each in the direction of the asymmetry 
feature and in the corresponding regions in quadrant IV.  

In  \S 2 and \S 3 we briefly describe the observations and data reduction. 
In \S 4, we discuss the velocities, spectral classification, metallicity 
estimates, and evidence for a population dependent asymmetry. In the final section 
we summarize the results of our kinematics analysis and its implications for the origin 
of the observed asymmetry.

\section{The Observations and Field and Object Selection}

We used HYDRA, a multi-object spectrometer on the Blanco 4--meter at Cerro 
Tololo Interamerican Observatory (CTIO), and on the WIYN\footnote{The WIYN Observatory is a 
joint facility of the University of Wisconsin-Madison, Indiana University, Yale 
University, and the National Optical Astronomy Observatory.} 3.5--meter at 
Kitt Peak National Observatory (KPNO). Each HYDRA instrument has two sets of fiber bundles.
WIYN-HYDRA has red and blue bundles with a 60$\arcmin$ field-of-view,
and we used the 3$\farcs$1 diameter blue fibers which cover wavelengths from
3000-7000\AA.  CTIO-HYDRA is an upgrade from WIYN-HYDRA with
40 additional fibers and faster fiber configuration.  CTIO-HYDRA
has a 40$\arcmin$ field of view and two fiber bundles, one  with 1$\farcs$3
apertures and a large bundle with 2$\farcs$0 apertures.  Unlike WIYN-HYDRA, 
both fiber bundles cover the spectral range 3000-11,000\AA.  CTIO-HYDRA has the
advantage of smaller fibers for more precise positioning and to potentially acquire 
dimmer objects. At the time of our observations, CTIO-HYDRA had just been 
commissioned and had not yet been upgraded to the 400~mm Bench Schmidt camera 
with a SiTe 2 $\times$ 4K array, but instead was used with the 229mm Air Schmidt 
camera with a Loral 1 $\times$ 3K engineering array.  This configuration prevented 
the use of all 138 large fibers, and to prevent the spectra from overlapping, we 
used every other fiber. Table \ref{tbl-01} summarizes the
characteristics of the CTIO and KPNO HYDRA spectrometers. Both instrument configurations
had a resolution of 1{\AA} per pixel.  Further information 
regarding the specifics of the instruments can be found at http://www.noao.edu/ctio/ 
or http://www.noao.edu/kpno/wiyn/.

We selected stars from four fields with a strong star count excess and two transition fields in quadrant I 
(Paper I) and their complementary matched fields in {\it l} and {\it b} in quadrant IV. Figure \ref
{fig-01} shows the distribution of the POSS I fields observed and at which 
observatory the observations were made. Using the Minnesota APS Catalog of the POSS I 
(http://aps.umn.edu), we randomly selected  stars between O magnitude 16 and 18 with colors between the 
peak of the color--magnitude distribution, $(O-E)_{peak}$ near $(O-E) = 1.0$ mag, 
and [$(O-E)_{peak}$ - 1] mag\footnote{On the POSS I photometric system
this color range, $(O-E)$ $\sim$ 0.0 to 1.0 mag., corresponds approximately 
to a $(B-V)$ $\sim$ 0.0 to 0.6, and includes stars in the blue and intermediate
color bins as defined in Paper I.}.  The contribution of the halo and thick disk stars 
to the star counts is expected to be more than 80\% (Paper I) in this magnitude and color range. 
For the CTIO fields, the 40$\arcmin$ diameter 
central region of each plate was queried. For the WIYN fields we were able to 
use a slightly larger 60$\arcmin$  diameter region. Each star was checked against 
known catalogs to eliminate any high proper motion stars. To prepare for the 
observations, we used a simulator to manually place each usable fiber on a star, 
a guide star or a blank sky field.  The instrument requires that there be 4 to 5 
FOPS stars, or bright guide stars of $\approx$ 10th magnitude, and for WIYN-HYDRA 
these stars had to be located near the edges of the field.  As a result of the 
precise APS astrometry, we were  able to easily acquire every guide star and  
the target stars with better than 95\% success rate in each field.

Table \ref{tbl-02} lists the APS plate number for each field, the 
galactic coordinates for the centers of the Hydra fields, the number of fibers 
dedicated to stars and to blank 
sky, the exposure times, and which night each field was observed.  For stars between
16-18th magnitude, long integrations were needed, and we used  3--4 exposures
for an average of 40 minutes each.  We also took a projector flat or quartz flat, 
and a comparison spectrum for each fiber configuration.  At CTIO a projector flat 
and comparison spectra were taken after each
exposure of the target stellar field. Because each fiber has a slightly different
response depending on its location,  the projector flats are used to remove any 
signature from the fiber itself.  At KPNO, projector flats and comparison spectra
were taken before a set of exposures on the initial field and at the end of the
exposures on the final field.  WIYN-HYDRA has a slower fiber positioner and these
procedures maximized observation time.  Comparison spectra were obtained and used to
dispersion correct the spectra.  We used the He-Ne-Ar lamp at CTIO, and the Cu-Ar
lamp at KPNO. Spectra of the  standards stars were observed throughout the night in the 
same manner using a single fiber manually positioned at the center of the field. 
Some stars observed on the first run, CTIO 1999, were repeated on the second run, 
CTIO 2000, as a further quality check.  Field P566 was observed on all three runs to 
compare results between KPNO and CTIO.

\section{Data Reduction and Velocity Measurement}

HYDRA has its own software reduction task DOHYDRA within the larger Image Reduction 
and Analysis Facility software package (IRAF\footnote{IRAF is written and supported by the IRAF programming group at the National Optical Astronomy Observatories (NOAO) in Tucson, Arizona. NOAO is operated by the  Association of Universities for Research in Astronomy (AURA), Inc. under cooperative agreement with the National Science Foundation}). DOHYDRA is a specialized package with 
multiple tasks for scattered light subtraction, flat fielding, fiber throughput 
correction, wavelength calibration, extraction, and sky subtraction.  There is a 
degree of automation and built-in record keeping that is necessary for the volume of 
data generated by this instrument. The task CCDPROC was used to correct each image for 
the overscan and bias.  Each individual exposure was processed with DOHYDRA using the 
projector flats for the flat field, the comparison spectra for the  dispersion correction, 
and a simple text table generated by the instrument to identify the apertures used.  
DOHYDRA first identifies the apertures, then flat fields the images by extracting 
and averaging the flat field spectra over all fibers.  Interactively, the average 
flat field is fit by a high order function, spline3 order 100, that normalizes the 
individual flat field spectra. The spectra were then dispersion corrected after identifying 
lines in the comparison spectra. The three 40 minute exposures were co-added and 
sky subtracted, and any remaining cosmic rays or bad pixels previously not eliminated 
were edited out. The spectra were not continuum subtracted as this made no difference 
for measuring the Doppler velocities. 

Nearly 1000 spectra were reduced using the above techniques.  Of
these 1000 spectra about 10\% were not used most often because the object 
was fainter than 17.5 magnitude, which turned out to be the practical limit for
adequate S/N for line identification and the velocity measurements. The radial
velocities were measured with the IRAF task RVIDLINES.  As a baseline, we manually
identified 2 to 3 spectral lines and then relied on RVIDLINES to automatically
identify the remaining lines from a list we created including the  Balmer series,
the Ca II  H and K lines, and strong metallic lines found in A-G-type dwarf
stars. We selected lines that could be clearly distinguished from the noise. 
Misidentified lines were rejected based on large velocity residuals. For the majority
of the stars the velocity error is better than $\sigma_{helio}$ $\sim$ 10 km s$^{-1}$.
Each night, standard stars were observed and reduced following the same procedures.
Table \ref{tbl-03} shows the  measured radial velocities of the standards and their
errors compared with their known velocities. The velocities are consistent for
all the nights and agree well with the published values.  In addition, we
independently compared the velocities for stars observed on multiple observing
runs.  Minor differences exist but were within the errors expected for
measurements at this resolution.  The duplicate observations were combined
by taking the weighted average radial velocity for each star. Accounting for
duplicates we have  radial velocities for 741 stars,  418 stars in quadrant I and 
323 stars in quadrant IV.  The heliocentric velocities are given in Table \ref{tbl-04}.

\section{Velocities, Spectral Classification and Metallicities}

In this section we review the measured velocities corrected to the Local Standard
of Rest (LSR) and the corresponding angular rotation rate ($\omega$) in the two
quadrants,  compare the stars' kinematics with their spectral types and metallicities,  
and discuss the kinematics of the three primary population groups in the two quadrants.

\subsection{The Kinematics} \label{lsrvel}

In  Figure \ref{fig-02}, we show the normalized histograms of the measured  velocities
for the quadrant I and IV stars corrected to the LSR using the solar motion values from
Hipparcos data reported in \cite{iba97}. Both quadrants show a very broad range
in velocities with high velocity tails to negative velocities in quadrant I and to positive
velocities in quadrant IV as we would expect,  but surprisingly both show net negative
velocities. After removing the obviously high velocity Population II stars
with V$_{LSR}$ greater than $\pm$ 200 km s$^{-1}$, the mean LSR--corrected  velocities
are -19.3 $\pm$ 3.2 km s$^{-1}$ and -15.3 $\pm$ 4.0 km s$^{-1}$ for quadrants I
and IV, respectively with standard deviations ($\sigma$) or dispersions of 65.3 km s$^{-1}$ 
and 68.8 km s$^{-1}$, respectively. If we assume a uniform,
axisymmetric thick disk, that is rotating with a circular velocity comparable
to the thin disk, an observer at the LSR would expect to measure stars with positive LSR
velocities in quadrant I, as these stars are moving away from us, and negative velocities
in quadrant IV since the stars would be approaching. The velocities should be comparable for
stars at similar distances and symmetric directions but of opposite sign.  However,
the thick disk is known to rotate slower or lag the thin disk by 30 -- 50 km s$^{-1}$.
Thus, the observed LSR velocities for the thick disk would be expected to show a
relative net shift with respect to the disk velocities in both quadrants, to more 
negative velocities in quadrant I and to more positive velocities in quadrant IV. 
Of course the velocity distributions in  Figure \ref{fig-02}
represent a mixture of disk, thick disk and halo stars, but the much  
greater shift with respect to the expected disk velocities of $\approx$ 10 to 25 km s$^{-1}$ in 
quadrant I, shows a clear asymmetry between the stars in the two quadrants and  suggests the 
presence of a population of stars with different kinematics
and greater net ``lag'' in quadrant I compared to quadrant IV. 

To investigate this possible lag or slower rotation in quadrant I, we have also calculated the
angular rate of rotation for each star, $\omega_{\star}$, using the standard equation,
\begin{equation}
V_{LSR} = (\omega_{\star} - \omega_{\odot})~ R_{\odot}~sin{\it l}~cos{\it b}
\end{equation}
with $R_{\odot}$ = 8 kpc (see \citet{rei98}), $V_{\odot}$ = 220 km s$^{-1}$, and $\omega_{\odot }$
= 27.5 km s$^{-1}$ kpc$^{-1}$.

The histograms for the angular rate of rotation in Figure \ref{fig-03}  show a
shift in the observed rotational rate and different kinematics between the stars in quadrants I
and IV. In their directions, and at 1 to 2 kpc from the Sun, these stars will typically be  6.4-7.5 
kpc from the Galactic Center. Using the power law fit to the rotation curve found in \citet{bb93}, 
stars at these galactocentric distances would have an expected $\omega$ of  
30 to 38 km s$^{-1}$ kpc$^{-1}$. Removing the high velocity  Population II stars with 
V$_{LSR}$ greater than $\pm$ 200 km s$^{-1}$, the mean $\omega_{\star}$ for quadrant IV 
is 33.4 $\pm$ 1.4 km s$^{-1}$ kpc$^{-1}$  consistent with circular rotation.  
However, the stars in quadrant I have a slower effective rotation, with a mean 
$\omega_{\star}$ = 21.8 $\pm$ 1.2 km s$^{-1}$ kpc$^{-1}$.  
{\it The results for V$_{LSR}$ and $\omega$ show a kinematic asymmetry 
between quadrants I and IV corresponding to the observed asymmetry in the star counts.}
However these results alone do not allow us to distinguish between radial streaming along
the major axis of a triaxial thick disk and a rotational lag due to interaction with
the bar in the disk. 

Furthermore, when we inspect the results for the individual fields (Table 5), we note that
there is a very interesting gradient or dependence of the mean V$_{LSR}$ values on longitude,
especially in quadrant I. The mean  V$_{LSR}$ values  become more negative
at the greater longitudes in quadrant I and somewhat more positive in quadrant IV. This 
dependence may let us  separate the radial and
tangential components to determine  if the kinematic asymmetry is due to streaming in
the radial direction or a lag entirely in the direction of Galactic rotation.

In \S 4.4 below, we use  additional information on the stars' metallicities to re-examine 
their  kinematics
as a function of stellar population and confirm that a significant fraction of the quadrant
I thick disk stars  are participating in a slower effective rotation . 

\subsection{The Spectral Types}

We have also determined approximate spectral types for all of the stars for which
we have a measured radial velocity. We call our types ``approximate'' because
we did not have observed standards for each spectral subtype, and because few standards
exist for the metal-weak population. Instead,
we used the reference table and online spectral catalog by \citet{j96} 
(ftp://ftp.noao.edu/catalogs/coudelib), plus spectra of our velocity standards 
in the classification process.  Comparison of the spectral types for stars 
observed at both CTIO '99 and WIYN '00 demonstrates that our method is internally consistent. However,
we note that the stars observed during the CTIO '00 run were subject to a small
light leak which produced an  offset in the flux level of each spectrum. 
This did not affect the velocity measurements which were confirmed by comparing
the results for the multiply observed stars. We classified these spectra according 
to the same criteria as the other stars, and comparison with the types for the stars 
with duplicate spectra show that the spectral types are consistent; although the classifications for 
these stars perhaps should be considered somewhat less reliable. All of the stars with 
spectra with sufficient S/N for the velocity measurements and spectral classification are
listed in Table \ref{tbl-04}.

Figure \ref{fig-04} shows a histogram of the spectral types for stars in quadrant
I compared to quadrant IV. There are approximately the same relative numbers of the
different spectral types in the two quadrants with slightly more early-type stars
(A8-F5) classified in quadrant I than in quadrant IV.  Comparing the spectral types
with the LSR velocities in Figure \ref{fig-05}, we note that more of the A8-G0 stars in quadrant I
have larger, in this case, more negative,  LSR velocities, suggestive of thick disk/inner
halo star velocities, than observed for the same types of stars in quadrant IV.

\subsection{Metallicities}

We obtained metallicity estimates for our stars using the methodology and  
calibration of \citet{b99}, which is based on the relationship between a
CaII K-line index, KP, and  the $(B-V)_0$ colors. The line indices KP and HP2 were
obtained using their procedures, but since we do not have
measured $(B-V)_0$ colors for our program stars, we have followed the procedure
described by them based on an estimate of the de-reddened color, from the  
Balmer-line index, HP2.  Previous experience suggests that this procedure
works quite well for stars of ``intermediate'' color, e.g., stars with $0.40 \le
(B-V)_0 \le 0.80$, with an estimated scatter on the order of 0.05 mags, but
exhibits somewhat higher scatter for stars with colors outside this range.
The great majority of our program stars fall within this optimal color range.  

Using this method we estimated the   metallicities for 581 of the 741 stars in Table 4.  
Metallicities were not measured from those spectra with the small light leak mentioned 
in  \S 4.2. Some of the spectra were also too noisy to obtain a reliable measurement 
and are so designated in the table. The normalized histograms
for the [Fe/H] values are shown in Figure \ref{fig-06}; the mean error in the individual 
[Fe/H] values ($\sigma_{[Fe/H]}$) is 0.25 dex. The distribution of the metallicities confirm our
selection criteria with a predominance of thick disk and halo stars in this color range. Removing 
the extremely metal poor stars with [Fe/H] $<$ -2.2, the stars in quadrant I have a mean 
[Fe/H] = -1.06 $\pm$ 0.07, and therefore appear to be slightly more metal rich  
than those in quadrant IV with a mean [Fe/H] = -1.27 $\pm$ 0.12.  Quadrant IV's mean [Fe/H]
is close what we would expect for a mixed metal-poor population.  Assuming that the stars 
in both quadrants are a mixed population, quadrant I apparently has relatively more stars 
with a higher 
metallicity, perhaps more representative of the thick disk population which has a
mean [Fe/H] of $\approx$ -0.7. This is consistent with the excess of presumed thick 
disk stars (Paper I) in the quadrant I fields yielding a slightly higher metallicity. 

\subsection{Population Separation and the Kinematics}

With this additional metallicity information, we use  [Fe/H] and $V_{LSR}$ to 
separate our sample into the three primary population groups and repeat our calculations of 
the mean $V_{LSR}$,  and $\omega$.  As in \S 4.1 and \S 4.3 we initially remove the 
extreme high velocity 
stars with $|V_{LSR}| \geq 200$ km s$^{-1}$ and define the three groups according to
metallicity; disk:  $-0.3 \le [Fe/H] \leq 0.30$, thick disk:  $-1.2 \le [Fe/H] \leq -0.30$
and halo:  $-2.2 \le [Fe/H] \leq -1.2$. Admittedly, given the uncertainties with our method 
for determining the metallicities (due primarily to the need for estimated colors), the errors 
in the individual [Fe/H] values, plus the natural
spread and overlap in [Fe/H] for these three groups, we do not expect a clean separation.
There is undoubtedly some contamination and overlap, especially between the disk and thick
disk stars given our criteria. For that we reason we have used the $V_{LSR}$ to further 
refine the separation. For example, a star  with an [Fe/H] indicative of 
the disk or thick disk but with a  $|V_{LSR}| \geq 150$ km s$^{-1}$,  was  reassigned to the 
halo population.  Figures \ref{fig-07} 
and \ref{fig-08} show the resulting histograms for $V_{LSR}$ and $\omega$, respectively for the  
halo, thick disk and disk populations in quadrants I and IV. The 
mean values with their errors and standard deviations are given in Table \ref{tbl-06}.  

The results for the presumed halo population are much as we would expect for a
population with a high velocity dispersion with respect to the LSR. With the large 
dispersion  in  their  velocities, the results for the halo population in the two 
quadrants do not appear to  be significantly different. The quadrant I halo stars,
however, do have a  slightly lower mean rotational rate than for those in quadrant IV.

The differences in the  mean velocities and rotation rates between the two quadrants for 
the thick disk and  disk stars confirm 
our previous  conclusion  that a significant population of quadrant I stars are 
rotating slower than those in quadrant IV.  Recall that the expected $V_{LSR}$ velocities 
are positive and negative in quadrants I and IV, respectively, and at the distances and 
directions of these stars should be on the order of 10 to 25 km s$^{-1}$.
The mean $V_{LSR}$ values for the presumed disk and thick disk stars in quadrant IV are
consistent with this expectation, but the thick disk stars do not show a significant positive
shift in $V_{LSR}$  or the slower effective rotation expected for this population. 
In quadrant I the shift to more negative velocities and the rotational lag is quite apparent
confirming the kinematic
asymmetry affecting the thick disk stars, and  to some extent the disk stars,  between 
quadrants I and IV. 
We also note that the velocity dispersions for the thick disk stars in both quadrants and 
for the disk stars in quadrant I  are
noticeably higher than the nominal dispersions for the thick disk ($\approx$ 45 km s$^{-1}$) and for 
main sequence disk G stars ($\approx$ 30 km s$^{-1}$) which may be due to a mixture from other
more high velocity populations in each case.  

In \S 4.1, we commented on the evidence for a gradient in the mean $V_{LSR}$ velocities
with Galactic longitude (see Table 4). Using  the additional information from the metallicities,
we have investigated the results for $V_{LSR}$ and $\omega$ as a function of population type
in the different fields; summarized in Table 7. The thick disk stars, and to a somewhat 
lesser extent, the halo stars in quadrant I\footnote{The disk stars are not included because 
there are too few  in the individual fields to yield meaningful results}, show a 
dependence of their  mean $V_{LSR}$ values on  longitude, while the  quadrant IV stars 
do not show a similar dependence. The observed $V_{LSR}$ is a combination of the  
stars' motions in the radial($v_{r}$), tangential($v_{\phi}$), and vertical($v_{z}$) 
directions (see eqn. 2) and the contribution of each to the total velocity will depend on 
the star's direction and distance . For example, looking  toward the Galactic center,  
P566 and P505 ({\it l} $\approx 20\arcdeg$), $v_{r}$ will be the primary component,   
while towards {\it l} of 40$\arcdeg$ to 50$\arcdeg$ the
radial and tangential contributions should be more nearly equal. Assuming that we have a uniform
population of stars in these different fields with a common or shared motion,  the observed 
dependence on longitude for the quadrant I stars should allow us to estimate the three components 
and determine if the quadrant I thick disk kinematics are due primarily to radial streaming or  
rotational lag. We have used the method of least squares applied to  
\begin{equation}
V_{LSR} = [v_{r}cos\theta + (\frac{v_{\phi}}{R_{\ast}} - \frac{V_{\odot}}{R_{\odot}})sin\theta]cosb + v_{z}sinb
\end{equation}
where  $\theta$ is the angle between the line of sight from the Sun to the star and the Galactic Center
to the star, $sin\theta = R_{\odot}{\sin}l/R_{\ast}$, and R$_{\ast}$ is the distance
of the star from the Galactic Center. Assuming the stars are 1 and 2 kpc from the Sun, 
we solved for $v_{r}$, $v_{\phi}$ and $v_{z}$.  
 The results are summarized in Table 8.  Not surprisingly, the uncertainty in 
the solutions is large given the small number of stars and the intrinsic dispersions of the thick
disk stars in these three directions. Nevertheless, the three parameter solution  for the quadrant I 
stars, shows  that
the slower effective rotation for these stars is due to slower motion in the direction of Galactic
rotation ($v_{\phi}$) with a lag of $\sim$ 80 km s$^{-1}$ with respect to circular rotation in the 
disk, and slower than the 40 -- 50  km s$^{-1}$ normally assumed for the thick disk.  The radial 
component shows a net motion 
toward the Galactic Center, but it is small and probably not significant.  
We repeated the same solutions for the quadrant IV stars. Although there are only 55 stars  
in this set, the results are distinctly different than those for quadrant I. The results for 
$v_{z}$ are sufficiently anomalous implying a large motion toward the plane and with a large error,
that it is probably meaningless. We therefore repeated the solutions for both quadrants with only
two parameters, setting $v_{z} = 0$. These solutions for quadrant IV for the two assumed distances
from the Sun, yielded more uniform results with smaller errors, and show only  a small rotational
lag  and a marginal radial outflow which is probably not 
significant. Interestingly, the quadrant I solutions for $v_{r}$ and $v_{\phi}$ are compatible with the
three parameter results with smaller errors. We conclude that the quadrant I thick disk stars
have a lag of 80 -- 90 km s$^{-1}$ in the direction of Galactic rotation while a similar
population in quadrant IV has a probable lag of only $\sim$ 20 km s$^{-1}$, confirming the 
kinematic asymmetry between the two directions.   

\section{Discussion -- Possible Causes of the Asymmetry}

In Paper I, we considered three possible explanations for the observed asymmetry
in the stars counts, a fossil remnant from an ancient merger, a triaxial thick
disk, and  an interaction with the bar in the old or thin disk in quadrant I, and
concluded that either the triaxial thick disk or a bar/thick disk interaction could
best explain the excess in the star counts, but that the merger remnant was less likely.
\citet{her92} and \citet{dbs98} showed that a rotating bar in the disk could induce
a gravitational ``wake'' that would trap and pile up stars behind it and these stars
may thus show a measurable lag in their rotational velocities. Likewise a triaxial thick 
disk could also yield different effective rotation rates because of non-circular streaming
motions along the major axis. 

Numerical simulations suggest that triaxial dark halos are a natural
consequence of Cold Dark Matter scenarios \citep{dub91}. Based on studies of rotating 
triaxial spheroids \citet{bs91} suggested that there may be both a rounder outer halo 
and a triaxial (flattened) inner halo. This model is supported by observations that 
show both a round outer halo (\citet{har87},\citet{cb00}) and evidence for a large flattened 
distribution in the inner halo (\citet{WysG89}, \citet{lh94}, \citet{cb00}).  \citet{cur99} 
also find a high 
probability that a rotationally supported disk and a non-axisymmetric halo can 
trigger instabilities leading to the formation of a long-lived bar.  Their 
simulations also suggest that these instabilities are similar in character to 
spiral-density waves, and can affect stars at higher Galactic latitudes in a 
similar way.  

For some time now, there has been substantial evidence for a bar within the bulge of
the Milky Way, but numerous more recent studies in the Galactic plane ($b$ $\pm$ 3$^{\circ}$)
 \citep{bs91,wei92,loc99,fw00,ham00} have revealed the presence of a stellar bar extending beyond the
bulge with a 3-5 kpc radius.  \citet{wei94} demonstrated that a structure with a rotating 
pattern speed in a disk system, assuming nearly circular orbits, generates three 
resonances: the inner Linblad (ILR), the outer Linblad (OLR) and another at the 
corotation location (CR), which can lead to significant kinematic signatures in the 
line-of-sight velocities and velocity distributions. In his model, a triaxial rotating 
spheroid will produce an increased velocity dispersion near the OLR at approximately 
4-5 kpc from the galactic center. The high overall velocity dispersion   
measured for the stars in our study may be explained by the OLR of a triaxial thick 
disk or inner halo. However, it is just as likely that losses in angular momentum 
of a moderate stellar bar may reduce the bar's pattern speed and inflate the velocity 
dispersions in the thick disk.  \citet{wei94} finds that the latter explanation 
supports his predictions of a stellar bar at 5 kpc with a position angle of 36$^{\circ}$, 
see Figure 3 in \citet{wei92}

The effect of a bar on its parent galaxy is dependent on the amount and distribution of
dark matter in the galaxy.  It has not yet been determined
if a stellar bar is created by the interaction between a thin disk and an outer (potentially
dark matter) halo, which then creates an asymmetric, triaxial thick disk/inner halo, or
whether, alternatively,  the presence of a triaxial thick disk/inner halo contributes 
to bar formation. In either case, the kinematics would be similar and the bar and major 
axis of the thick disk would be  nearly aligned.  If the leading edge of the triaxial thick
disk is in quadrant I, as implied by the star count excess, the differences between the  rotational 
kinematics would lead to a more negative lag in the velocities in quadrant I as would a gravitational 
interaction with the disk bar also in quadrant I. Thus  our results for a kinematic asymmetry 
between quadrants I and IV and slower effective rotation in quadrant I could be due to either cause.

The mean $V_{LSR}$ velocities and rotation rates ($\omega$) for the separate fields 
in quadrant I also show a  dependence on Galactic longitude for the thick stars not shown
by the stars in quadrant IV. Least squares  solutions for the three velocity components
of the $V_{LSR}$ show that the thick disk stars in quadrant I  have  a major lag in their
velocities in the direction of Galactic rotation  much slower than
for the quadrant IV stars and slower than the nominal thick disk lag, with no significant 
radial component. This result supports a gravitational  interaction with the bar as the
more probable explanation for the observed asymmetries. This  does not
necessarily mean that the thick disk is not triaxial. As discussed above, 
bar/thick disk interactions may be independent of a triaxial thick disk, or it might be the case that
triaxial thick disks are a dynamical result of bar formation or vice versa and the two may be
nearly aligned.

To further investigate the possibility that the asymmetry feature is a consequence of
a gravitational wake induced by a stellar bar,
Debattista and Sellwood (private communication) have added a thick disk component to their models, and 
followed the evolution as a bar forms and rotates in the thin disk.  Their preliminary results
show that the non-axisymmetric response in the thick disk induced by the
interaction with the stellar bar can give better agreement with the
observed asymmetries in the star counts and kinematics.  These models
and  a detailed comparison with the observations will be presented in a later paper
\citep{dbs03}. 

In Paper I, we found a significant excess of thick disk/inner halo stars in
quadrant I above and below the galactic plane extending from $l$ $\sim$ 20$^\circ$
-- 55$^\circ$ and $|b|$ 20$^\circ$ -- 45$^\circ$ covering more than 600 ($\times$ 2) square 
degrees.  Doppler velocities, spectral types and metallicities reported here from 
spectra of 741 stars in quadrants  I and IV reveal a kinematic asymmetry as well which
shows most strongly among the thick disk stars.  The stars in quadrant I 
have a negative  mean LSR velocity with a significant net shift in the expected
LSR velocity, a  slower effective rotation rate, and a slightly higher mean metallicity 
than the stars in quadrant IV. {\it  We also demonstrate that the quadrant I thick disk stars 
have a significant lag in their velocities in the direction of galactic 
rotation,  much slower than found for the quadrant IV stars and slower than normally assumed.} 
These results tend to support a gravitational interaction between the thick disk stars 
and the rotating stellar bar in the disk as the best explanation for the asymmetry in 
the star counts and the kinematics. Combined with the evidence reported by \citet{lh03} 
that some  parameters such as the scale height and the normalization may vary with direction, 
the characteristics of the thick disk population, including its kinematics, may depend 
on where one is looking.

After this paper was submitted, the discovery of the  Canis Major merger remnant 
was announced by  \citet{Mar04}. Their detection of  an extensive remnant relatively 
nearby close to the Galactic plane,  suggests that such events may be the origin of the thick disk. 
In Paper I we rejected a merger remnant as an explanation for the star count excess
in quadrant I  for several reasons, including the shape and extent of the asymmetry
region on the sky, its presence above and below the plane, and a comparison with the
debris field of the Sagittarius dwarf merger. 
However, the debris left by Canis Major or other mergers could contribute to
our observed  asymmetry in the star counts and kinematics and to the
varying  characteristics of the thick disk with direction \citep{lh03}. 
The region of our observed asymmetry is not reproduced in the simulations for 
Canis Major \citep{Mar04}, although this does not necessarily eliminate the role 
of this or another merger.
Indeed the star count excess and much slower rate of rotation for the thick disk stars
in the asymmetry feature could be due to interaction of a merger remnant with the bar in
the disk. An important question to be decided with additional observations and modelling 
is whether the thick disk/inner halo populations sampled in quadrants I and IV are a 
single population with asymmetric spatial and kinematic properties or are two different
populations.  
 
Future work will include additional velocity measurements at higher longitudes and in
fields below the Galactic plane plus the addition of proper motions from the APS Catalog
and other all--sky surveys. These additional observations will help to further clarify
the origin of the asymmetry.
\acknowledgments

We are grateful to the support staff at both CTIO and KPNO for their
excellent support with the HYDRA spectrograph and we especially want to
thank Nick Suntzeff and Riccardo Venegas at CTIO.
We are pleased to acknowledge very useful commuications with Jerry Sellwood and
Victor Debattista. We also thank Kris Davidson for his programming assistance
with the least squares solutions. 
The APS project has been supported by the NSF, NASA and the
University of Minnesota. This work was supported in part by NASA's
Applied Information Systems Research Program (NAG 5-9407).
T.C.B. acknowledges partial support for this work from NSF grants AST 00-98508 and
AST 00-98549, awarded to Michigan State University.


\clearpage




\figcaption[JParker_fig01.eps]{A map of the selected fields observed at CTIO and KPNO. The light gray identifies observations made on the Blanco 4m at CTIO in 1999, dark gray for the 2000 observations and black indicates those fields observed with the WIYN 3.5m at KPNO in 2000. \label{fig-01}}

\figcaption[JParker_fig02.eps]{Normalized histograms of the LSR velocities for quadrants I(a) and IV(b). \label{fig-02}}

\figcaption[JParker_fig03.eps]{Normalized histograms of the angular rotation rates between quadrants I(a) and IV(b). \label{fig-03}}

\figcaption[JParker_fig04.eps]{A normalized histogram of the spectral types for quadrants I and IV. \label{fig-04}}

\figcaption[JParker_fig05.eps]{LSR velocities versus spectral types for quadrant I(a) and IV(b). \label{fig-05}}

\figcaption[JParker_fig06.eps]{Normalized histograms of metallicities for quadrants I and IV. a) Q1 has an average metallicity [Fe/H] = -1.06 $\pm$ 0.07 whereas b) quadrant IV has a mean [Fe/H] = -1.27 $\pm$ 0.12.  \label{fig-06}}

\clearpage

\figcaption[JParker_fig07.eps]{The distribution  of  $V_{LSR}$ with population type in  quadrants I and IV;  a-b) halo c-d) thick disk e-f) disk. \label{fig-07}}

\figcaption[JParker_fig08.eps]{The distribution  of  $\omega$ with population type in  quadrants I and IV; a-b) halo c-d) thick disk e-f) disk. \label{fig-08}}


\clearpage
\begin{deluxetable}{llccc}
\tabletypesize{\scriptsize}
\tablewidth{0pt}
\tablecaption{HYDRA: Instrument Specifics \label{tbl-01}}
\tablehead{
\colhead{Parameter}  & \colhead{Units}   &\multicolumn{3}{c}{Observing Runs} \\
\cline{3-5}
    &         & \colhead{CTIO 4-m}     & \colhead{CTIO 4-m}   & \colhead{KPNO/WIYN 3.5-m}  \\
    &         & \colhead{Apr. 15-18, 1999}  & \colhead{May 27-30, 2000}  & 
\colhead{June 2-5, 2000}}
\startdata 
CCD        &                 & Loral 3K-1   & Loral 3K-1  & T2KC    \\
Grating    &                 & KPGL-1       & KPGL-1      & KPC007  \\
Grating Tilt & [$^{\circ}$]  & 7.923        & 7.923       & 23.047  \\
Gain       & [e$^{-}$/ADU]   & 1.99         & 1.99        & 1.7     \\
Lines/mm   &                 & 632          & 632         & 632     \\
Read Noise & [e$^{-}$/rms]   & 7.8          & 7.8         & 4.3     \\
Binning    &                 & 1$\times$2   & 1$\times$2  & 1$\times$1 \\
Central $\lambda$ & [\AA]    & 4167         & 4167        & 4270    \\
Dispersion & [\AA/pix]       & 1.0378       & 1.0378      & 0.986   \\
Fibers     &                 & Large        & Large       & Blue    \\
Usable Fibers &             & 70/138       & 70/138       & 98/100 \\
Quartz &                     & He-Ne-Ar     & He-Ne-Ar    & Cu-Ar   \\
\enddata
\end{deluxetable}

\clearpage
\begin{deluxetable}{lrrccll}
\tabletypesize{\scriptsize}
\tablewidth{0pt}
\tablecaption{Observations \label{tbl-02}}
\tablehead{
\colhead{Field}   & \colhead{$l$[$^{\circ}$]}  & \colhead{$b$ [$^{\circ}$]}
 & \multicolumn{2}{c}{Fibers} & \colhead{Exp Time}  & \colhead{Nights}  \\
\cline{4-5} \\
 &   &  & \colhead{Stars}  & \colhead{Sky} & \colhead{[min]}  & \colhead{Observed} }
\startdata
CTIO 1999 &      &       &        &                 &   \\[0.1in]
P913    & 320.0 & 30.0   & 54 & 4 & 40 $\times$ 4 & April 15 \\
P566    & 20.0  & 32.0   & 50 & 5 & 40 $\times$ 3, 60  & April 15,17 \\
P858    & 329.0 & 32.0   & 54 & 5 & 40 $\times$ 4 & April 16 \\
P802    & 339.5 & 33.0   & 55 & 5 & 40 $\times$ 4 & April 17 \\
P741    & 339.0 & 41.0   & 53 & 5 & 40 $\times$ 4 & April 17 \\
P799    & 320.0 & 41.0   & 52 & 4 & 40 $\times$ 3 & April 18 \\
P505    & 21.0  & 42.5   & 50 & 5 & 40 $\times$ 3 & April 18 \\
\hline                                                      \\
CTIO 2000 &      &       &        &                 &       \\[0.1in]
P913    & 320.0 & 30.0   & 56 & 5 & 40 $\times$ 3 & May 27     \\
P802    & 339.5 & 33.0   & 49 & 4 & 40 $\times$ 2 & May 27     \\
P566    & 20.0  & 32.0   & 57 & 3 & 40, 40, 40    & May 27,28,29 \\
P799    & 320.0 & 41.0   & 52 & 5 & 40 $\times$ 3 & May 28    \\
P858    & 329.0 & 32.0   & 50 & 5 & 40 $\times$ 3, 40  & May 28,29   \\
P855    & 309.0 & 37.0   & 55 & 4 & 40 $\times$ 3 & May 29     \\
P741    & 339.0 & 41.0   & 56 & 5 & 40 $\times$ 3 & May 29     \\
\hline                                                      \\
WIYN 2000 &      &       &        &                  &  \\[0.1in]
P387    & 40.0  & 41.0   & 80 & 5 & 40 $\times$ 3 & June 2 \\
P448    & 40.0  & 30.0   & 77 & 5 & 40 $\times$ 3 & June 2 \\
P505    & 21.0  & 42.5   & 81 & 5 & 40 $\times$ 3 & June 3 \\
P566    & 20.0  & 32.0   & 79 & 5 & 40 $\times$ 3 & June 3 \\
P507    & 31.0  & 32.0   & 83 & 5 & 40 $\times$ 3 & June 4\\
P332    & 51.0  & 37.0   & 85 & 5 & 40 $\times$ 3 & June 4 \\
P566-1  & 20.0  & 32.0   & 77 & 5 & 40 $\times$ 3 & June 5 \\
\enddata
\end{deluxetable}

\clearpage
\begin{deluxetable}{lrrrrrr}
\tabletypesize{\scriptsize}
\tablewidth{0pt}
\tablecaption{Radial Velocity Standards \label{tbl-03}}
\tablehead{
\colhead{Star}  & \colhead{$V_{r}$ \tablenotemark{a}}[km s$^{-1}$]  & \colhead{Spectral Type \tablenotemark{a}}
 & \multicolumn{4}{c}{Measured Radial Velocities [km s$^{-1}$]} \\[0.1in] \cline{4-7} \\
        &                 &               & \colhead{night 1} & \colhead{night 2} & \colhead{night 3}
         & \colhead{night4}}
\startdata
CTIO 1999 &                    &               & April 15       & April 16 
 & April 17 & April 18 \\[0.1in]
HD 136202  & 53.5 $\pm$ 0.2  & F8 IV-V         & 54.0 $\pm$ 4.0 &  52.7 $\pm$ 3.4        & 53.2 $\pm$ 4.3  & 53.3 $\pm$ 4.4 \\
HD 114762  & 49.9 $\pm$ 0.5  & dF7/F9 V        & -              &  47.5 $\pm$ 5.5        & 49.6 $\pm$ 6.6  & 49.5 $\pm$ 2.0 \\
HD 86986   & 8.7              & A1 V           & 8.3 $\pm$ 3.0  & 8.9 $\pm$ 2.6 & 8.9 $\pm$ 3.0  & 8.7 $\pm$ 2.2 \\
HD 89449   &    6.5 $\pm$ 0.5  & F6 IV         & -              & 6.7 $\pm$ 1.9 & 6.8 $\pm$ 2.5  & 6.3 $\pm$ 2.4 \\
\hline                                 \\
CTIO 2000 &                  &               & May 26 & May 27 & May 28 & May 29 \\[0.1in]
HD 180134  &  -23.5  & F7 V & - & -23.7 $\pm$ 5.9 & -22.9 $\pm$ 4.7 & -23.2 $\pm$ 4.2 \\
HD 188642  &  -30.8  & F4 V & - & -31.6 $\pm$ 3.8 & -30.2 $\pm$ 5.9 & -30.7 $\pm$ 2.4 \\
HD 203638  &   21.9  & KO III  & - & 21.3 $\pm$ 4.0 & 19.7 $\pm$ 3.1 & 20.8 $\pm$ 5.4 \\
\hline                                          \\
WIYN 2000 &                  &               & June 2 & June 3 & June 4 & June 5 \\[0.1in]
HD 136202  &   53.5 $\pm$ 0.2  & F8 IV-V &  52.5  $\pm$ 4.9 & 53.3 $\pm$ 4.9 & 53.2 $\pm$ 4.0 &  - \\
HD 114762  &   49.9 $\pm$ 0.5  & dF7/F9 V &  51.0 $\pm$ 4.9 & 50.8 $\pm$ 6 & 51.0 $\pm$ 5.5 & 51.9 $\pm$ 4.8 \\
HD 161817  & -362.8          & sdA2    & -367.4 $\pm$ 4.8 & -364.0 $\pm$ 3.8 & -364.5 $\pm$ 1.7 & -362.3 $\pm$ 6.5 \\
BD 28 3402 &  -36.6 $\pm$ 0.5  &  F5 V    &  -36.7 $\pm$ 3.4 & -36.1 $\pm$ 3.7 & -37.8 $\pm$ 2.5 & -37.8 $\pm$ 3.3 \\
\enddata
\tablenotetext{a}{\scriptsize standard values reported from SIMBAD http://simbad.u-strasbg.fr/Simbad}

\end{deluxetable}

\clearpage
\begin{deluxetable}{llllllrrclll}
\tabletypesize{\scriptsize}
\tablewidth{0pt}
\tablecaption{Velocities,  Spectral Types and Metallicities \tablenotemark{a} \label{tbl-04}}
\tablehead{
\colhead{Field \tablenotemark{b}} & \colhead{APS Star \tablenotemark{c}} 
& \colhead{$O$ Mag} & \colhead{$(O-E)$}  & \colhead{$l$[$^{\circ}$]} & \colhead{$b$[$^{\circ}$]}
& \colhead{$V_{helio}$} & \colhead{$\sigma_{helio}$} & \colhead{Number}
 & \colhead{[$\frac{Fe}{H}]$ \tablenotemark{d}}
& \colhead{$\sigma_{[\frac{Fe}{H}]}$} & \colhead{Spec.}  \\
\colhead{/Obs. Run}  & \colhead{Number}  &   &   &   &   & \colhead{km s$^{-1}$}  & \colhead{km s$^{-1}$}  &  \colhead
{Lines}  &  &  & \colhead{Type}} 
\startdata
P332/3	&	1931756	&	16.1	&	0.8	&	51.3	&	37.1	&	12.5	&	3.6	&	4	&	-0.71	&	0.15	&	G5-G8	\\	
	&	2058237	&	16.5	&	0.6	&	50.9	&	37.5	&	-80.3	&	11.5	&	4	&	 0.30	&	0.22	&	GO	\\	
	&	2500783	&	16.0	&	0.6	&	50.3	&	36.8	&	-31.8	&	 3.7	&	6	&	 0.30	&	0.15	&	G5	\\	
	&	2199563	&	16.1	&	0.9	&	50.5	&	37.5	&	-108.6	&	11.1	&	6	&	-0.99	&	0.31	&	G5-G8	\\	
	&	1922793	&	15.9	&	0.7	&	51.1	&	37.4	&	-84.8	&	 9.9	&	8	&	-0.46	&	0.35	&	G5-G8	\\	
	&	2215458	&	16.3	&	0.6	&	50.7	&	36.9	&	-153.4	&	 8.8	&	4	&	-1.01	&	0.15	&	F5	\\	
\enddata
\tablenotetext{a}{The complete version of this table is in the electronic edition of the Journal. 
The printed edition contains only a sample.}

\tablenotetext{b}{\scriptsize 1, 2, 3
refer to the three observing runs CTIO 1999, CTIO 2000, and WIYN 2000 respectively.}

\tablenotetext{c}{\scriptsize Positions for the stars can be obtained from the Minnesota Automated Plate Scanner(MAPS) Catalog of the POSS I using either the sql-based query for individual stars (http://aps.umn.edu/catalog/sql.html) or the skybox query ((http://aps.umn.edu/catalog) for a region on the sky.}

\tablenotetext{d}{\scriptsize 
Stars with spectra considered too noisy to obtain reasonable abundance estimates have a ... as their [Fe/H] value
and are not used in the measurement of the mean. 
Stars with inferred $(B-V)_0$ colors outside the ``optimum range'' are labeled with a ``:''.
Stars with a noisy spectrum, but have an abundance estimate are labeled with a ``N''.}

\end{deluxetable}

\clearpage
\begin{deluxetable}{llrcccc}
\tabletypesize{\scriptsize}
\tablewidth{0pt}
\tablecaption{Mean LSR Velocities and Rotation Rates for  Quadrant I and IV Fields \label{tbl-05}}
\tablehead{
\colhead{Field} & \colhead{{\it l, b}} & \colhead{$N_{stars}$} & \colhead{$<V_{LSR}>$}  & \colhead{$\sigma_{V_{LSR}}$} &  \colhead{$<\omega>$}  & \colhead{$\sigma_{\omega}$}\\
& & & \colhead{km s$^{-1}$} &  \colhead{km s$^{-1}$} &  \colhead{km s$^{-1}$kpc$^{-1}$} & \colhead{km s$^{-1}$kpc$^{-1}$} }
\startdata
Q I &  &  405 & -19.2 $\pm$ 3.2  & 62.3  & 21.8 $\pm$ 1.2 & 23.2\\[0.1in]
P332 &  51.0, 37.0 & 37 & -42.1 $\pm$ 8.4 & 51.2 &  19.0 $\pm$ 1.7 & 10.4\\
P387 &  40.0, 41.0 &  57 & -35.0 $\pm$ 7.6 & 57.3 & 18.4 $\pm$ 2.0 & 14.9\\
P448 &  40.0, 30.0 &  60 & -15.9 $\pm$ 8.1 & 62.8 & 23.9 $\pm$ 1.8 & 14.1\\
P507 &  31.0, 32.0 &  65 & -20.7 $\pm$ 8.6 & 69.5 & 21.6 $\pm$ 2.5 & 20.0\\
P505 &  21.0, 42.5 &  64 & -26.4 $\pm$ 8.3 & 66.2 & 14.9 $\pm$ 3.9 & 31.5\\
P566 &  20.0, 32.0 & 122 & -1.0  $\pm$ 5.7 & 63.2 & 27.1 $\pm$ 2.5 & 27.1\\ 

\hline\\ 
Q IV  &  & 317  & -15.3 $\pm$ 3.9  & 68.8  & 33.4 $\pm$ 1.4 & 24.2 \\[0.1in]
P855 &  309.0, 37.0 & 31 & 14.9 $\pm$ 10.8 & 60.2 & 24.5 $\pm$ 2.2 & 12.2\\
P799 &  320.0, 41.0 & 60 & -8.7 $\pm$ 8.5 & 66.1 & 29.7 $\pm$ 2.2 & 17.4\\
P913 &  320.0, 30.0 & 54 & -16.1 $\pm$ 7.9 & 58.0 & 31.1 $\pm$ 1.8 & 13.1\\
P858 &  329.0, 32.0 & 43 & -14.5 $\pm$ 13.1 & 86.3 & 31.7 $\pm$ 3.8 & 25.1\\
P741 &  339.0, 41.0 & 63 & -33.3 $\pm$ 8.3 & 65.5 & 42.7 $\pm$ 3.9 & 30.7\\
P802 &  339.5, 33.0 & 66 & -18.2 $\pm$ 8.8 & 71.8 & 35.1 $\pm$ 3.7 & 30.4\\
\enddata
\end{deluxetable}

\clearpage
\begin{deluxetable}{llcccc}
\tabletypesize{\scriptsize}
\tablewidth{0pt}
\tablecaption{Mean LSR Velocities and Rotation Rates for the Three Populations \label{tbl-06}}
\tablehead{
\colhead{Population} & \colhead{$N_{stars}$} & \colhead{$<V_{LSR}>$}  & \colhead{$\sigma_{V_{LSR}}$} &  \colhead{$<\omega>$}  & \colhead{$\sigma_{\omega}$}\\
& & \colhead{km s$^{-1}$} &  \colhead{km s$^{-1}$} &  \colhead{km s$^{-1}$kpc$^{-1}$} & \colhead{km s$^{-1}$kpc$^{-1}$} }
\startdata
Q I            &       &       &       &       &              \\[0.1in]
Disk              & 24    & -6.9  $\pm$ 9.8   &  47.8 & 26.4 $\pm$3.3 & 16.2    \\
Thick Disk    & 195   & -12.9 $\pm$ 4.1   &  57.6 & 24.5 $\pm$ 1.4  & 19.4    \\
Halo              & 152   & -34.4 $\pm$ 5.7  &  70.5 &  16.2 $\pm$ 2.1   & 25.9    \\
\hline                                                        \\
Q IV           &       &       &       &       &              \\[0.1in]
Disk              & 8     & -16.1 $\pm$ 12.6   & 35.9  & 30.7 $\pm$ 3.7  & 10.4     \\
Thick Disk    & 55    & -24.1 $\pm$ 8.4  & 62.3  & 36.4 $\pm$ 2.9 & 21.5    \\
Halo              & 99   & -27.8 $\pm$ 7.2   & 71.4  & 38.2 $\pm$ 2.7  & 27.0    \\
\enddata
\end{deluxetable}

\clearpage
\begin{deluxetable}{llrcccc}
\tabletypesize{\scriptsize}
\tablewidth{0pt}
\tablecaption{Mean LSR Velocities and Rotation Rates for the Populations in Different Fields \label{tbl-07}}
\tablehead{
\colhead{Field \& Population} & \colhead{{\it l, b}} &  \colhead{$N_{stars}$} & \colhead{$<V_{LSR}>$}  & \colhead{$\sigma_{V_{LSR}}$} &  \colhead{$<\omega>$}  & \colhead{$\sigma_{\omega}$}\\
& & & \colhead{km s$^{-1}$} &  \colhead{km s$^{-1}$} &  \colhead{km s$^{-1}$kpc$^{-1}$} & \colhead{km s$^{-1}$kpc$^{-1}$} }
\startdata
Q I     &       &       &       &       &       &              \\[0.1in]
P332 &  51.0, 37.0 &   &       &       &       &              \\
Thick Disk &   &   23  & -35.6 $\pm$ 9.2 & 44.1 & 20.3 $\pm$ 1.9 & 9.0\\
Halo       &   &   6   & -105.0 $\pm$ 22.4  & 54.9 &  6.2 $\pm$ 4.5 & 11.1\\[0.1in]

P387 &  40.0, 41.0 &   &       &       &       &              \\
Thick Disk &   &  22   & -22.4 $\pm$ 8.0 & 37.5 & 21.7 $\pm$ 2.1 & 9.7\\
Halo       &   &  31   & -48.7 $\pm$ 10.8 & 60.2 & 14.9 $\pm$ 2.8 & 15.6\\[0.1in]

P448 &  40.0, 30.0 &    &       &       &       &              \\
Thick Disk &   &  37 & -18.7 $\pm$ 8.4  & 51.1 & 23.3 $\pm$ 1.9 & 11.5 \\
Halo  &   &       15 & -33.0 $\pm$ 20.0 & 77.5 & 20.0 $\pm$ 4.5 & 17.4 \\[0.1in]

P507 &  31.0, 32.0 &    &       &       &       &              \\
Thick Disk &   &  44  & -14.3 $\pm$ 10.4 & 69.0 & 23.4 $\pm$ 3.0 & 19.8 \\
Halo     &   &    15  & -36.0 $\pm$ 20.8 & 80.5 & 17.2 $\pm$ 6.0 & 23.1 \\[0.1in] 

P505 &  21.0, 42.5  &    &       &       &       &              \\
Thick Disk &   &  21 &  -5.6 $\pm$ 10.6  & 48.4 & 24.8 $\pm$ 5.0 & 23.0\\
Halo     &   &    38 &  -31.0 $\pm$ 11.3 & 69.5 & 12.8 $\pm$ 5.3 & 32.9\\[0.1in]

P566 &  20.0, 32.0  &    &       &       &       &              \\
Thick Disk &   &  48 & +5.1 $\pm$ 8.9 & 61.9 & 29.7 $\pm$ 3.8 & 26.6 \\
Halo  &   &       47 & -18.7 $\pm$ 9.6 & 66.0 & 19.5 $\pm$ 41. & 28.4\\

\hline\\   

Q IV &       &       &       &       &       &              \\[0.1in]
P855\tablenotemark{a}  & 309.0, 37.0  &   &     &    &                \\

P799  & 320.0, 41.0 & &       &       &       &              \\
Thick Disk &   &  9 & -23.9 $\pm$ 23.6 & 70.7 & 33.7 $\pm$ 6.1  & 18.2\\
Halo  &   &       9 & -11.8 $\pm$ 20.5 & 61.4 & 30.6 $\pm$ 5.3  & 15.8\\[0.1in]

P913  & 320.0, 30.0& &       &       &       &              \\
Thick Disk &   &  10 & -18.5 $\pm$ 16.1 & 50.8 & 31.7 $\pm$ 3.6 & 11.4\\
Halo  &   &       18 & -29.5 $\pm$ 13.9 & 59.0 & 34.1 $\pm$ 3.1 & 13.3\\[0.1in]

P858 & 329.0, 32.0 & &       &       &       &              \\
Thick Disk &   &  6  & -20.3 $\pm$ 41.9 & 102.6 & 33.3 $\pm$ 12.1 & 29.7\\
Halo  &   &      24  & -8.9 $\pm$  14.5 & 71.1  & 30.0 $\pm$ 4.2  & 20.6\\[0.1in]

P741  & 339.0, 41.0  & &       &       &       &              \\
Thick Disk &   &  12 & -43.9 $\pm$ 12.2 & 42.4 & 47.4 $\pm$ 5.6 & 19.6\\
Halo  &   &       23 & -57.9 $\pm$ 13.3 & 64.0 & 54.1 $\pm$ 6.1 & 29.4 \\[0.1in]

P802  & 339.5, 33.0  & &       &       &       &              \\
Thick Disk &   &  18 & -23.0 $\pm$ 12.2 & 51.6 & 33.9 $\pm$ 5.3 & 22.6\\
Halo  &   &       24 & -21.6 $\pm$ 16.6 & 81.5 & 37.0 $\pm$ 6.9 & 33.9\\

\enddata
\tablenotetext{a}{No [Fe/H] measurements for population separation.}
\end{deluxetable}

\clearpage
\begin{deluxetable}{lccccc}
\tabletypesize{\scriptsize}
\tablewidth{0pt}
\tablecaption{Results of the Solutions for the Velocity Components of the Thick Disk Motions\label{tbl-08}}
\tablehead{
\colhead{Quadrant}  &  \colhead{$N_{stars}$} &  \colhead{Distance}  &\colhead{$v_{r}$} &  \colhead{$v_{\phi}$} &  \colhead{$v_{z}$}\\
& & &  \colhead{km s$^{-1}$} & \colhead{km s$^{-1}$} & \colhead{km s$^{-1}$}
}
\startdata
Q I   &  195 &  1 kpc & -40 $\pm$ 26   &  146 $\pm$ 28 &  -23 $\pm$ 26 \\
&  &            2 kpc & -33 $\pm$ 23   &  141 $\pm$ 31 &  -18 $\pm$ 46 \\[0.1in]

Q I   &  195 &  1 kpc &  -28 $\pm$ 15  &  136 $\pm$ 21 &   ... \\
&  &            2 kpc &  -23 $\pm$ 15  &  132 $\pm$ 18 &   ... \\[0.1in]

\hline\\

Q IV &  55  &   1 kpc & -38 $\pm$ 57   &  173 $\pm$ 42 &  -107 $\pm$ 87 \\
&  &            2 kpc &  -45 $\pm$ 71  &  157 $\pm$ 39 &  -113 $\pm$ 88 \\[0.1in]

Q IV  & 55 &    1 kpc & 31 $\pm$ 34   &   207 $\pm$ 53 &  ... \\
&  &            2 kpc & 26 $\pm$ 34   &   197 $\pm$ 47 &  ...\\[0.1in]

\enddata
\end{deluxetable}

\end{document}